\documentclass[twocolumn, prl, aps, superscriptaddress]{revtex4-1}

\usepackage{graphicx}
\usepackage{amsmath}
\usepackage{bm}
\usepackage{xcolor}
\usepackage{color}
\usepackage{siunitx}

\newcommand{\beq}{\begin{equation}}
\newcommand{\eeq}{\end{equation}}
\newcommand{\bea}{\begin{eqnarray}}
\newcommand{\eea}{\end{eqnarray}}

\begin{document}

\title{Functional Form of the Superconducting Critical Temperature from Machine Learning}
\author{S.~R. Xie}
\affiliation{Department of Materials Science and Engineering,  University of Florida, Gainesville FL 32611, USA}
\affiliation{Quantum Theory Project, University of Florida, Gainesville FL 32611, USA}
\author{G.~R. Stewart}
\affiliation{Department of Physics,  University of Florida, Gainesville FL, 32611 USA}
\author{J.~J. Hamlin}
\affiliation{Department of Physics,  University of Florida, Gainesville FL, 32611 USA}
\author{P.~J. Hirschfeld}
\affiliation{Department of Physics,  University of Florida, Gainesville FL, 32611 USA}
\author{R.~G. Hennig}
\email{rhennig@ufl.edu}
\affiliation{Department of Materials Science and Engineering,  University of Florida, Gainesville FL 32611, USA}
\affiliation{Quantum Theory Project, University of Florida, Gainesville FL 32611, USA}
\date{\today}

\begin{abstract}
Predicting the critical temperature $T_c$ of new superconductors is a notoriously difficult task, even for electron-phonon paired superconductors for which the theory is relatively well understood. Early attempts to obtain a simple $T_c$ formula consistent with strong-coupling theory, by McMillan and Allen and
Dynes, led to closed-form approximate relations between $T_c$ and various measures of the phonon spectrum and the electron-phonon interaction appearing in Eliashberg theory.  Here we propose that these approaches can be improved with the use of machine learning algorithms. As an initial test, we train a  model for identifying low-dimensional descriptors using the $T_c < 10$~K data tested by Allen and Dynes, and show that a simple analytical expression thus obtained improves upon the Allen-Dynes fit.  Furthermore, the  prediction for the recently discovered high $T_c$ material H$_3$S at high pressure is quite reasonable.  Interestingly, $T_c$'s for more recently discovered superconducting systems with a more two-dimensional electron-phonon coupling, which do not follow Allen and Dynes' expression, also do not follow our analytic expression.  Thus, this machine learning approach appears to be a powerful method for highlighting the need for a new descriptor beyond those used by Allen and Dynes to describe their set of isotropic electron-phonon coupled superconductors. We argue that this machine learning method, and its implied need for a descriptor characterizing Fermi surface properties,  represents a promising new approach to superconductor materials discovery which may eventually replace the serendipitous discovery paradigm begun by Kamerlingh Onnes.     
\end{abstract}
\maketitle{}

\section{Introduction}
Discovery of new superconductors has historically proceeded largely serendipitously, with guidance from rules of thumb (such as Matthias' e/a ratio) rather than many-body and ab-initio theory.  The space of possible materials to search for new superconductors is vast, considering that many  discoveries in the last thirty years are multinary compounds.  Thus, it is desirable to appeal to recent computational developments, aided by theory, to assist this process.  The history of ab-initio and materials-genome type approaches to superconducting materials discovery has recently been reviewed by Norman~\cite{genome_SC},  Pickett~\cite{Pickett2018}, and Duan et al\cite{Duan_review2019}.

While initially, success in prediction (as opposed to analysis after discovery, {\it i.e.}, postdiction) was rare to nonexistent, more recently the potential for theory to aid in the discovery of new high-temperature superconductors was  dramatically demonstrated by the prediction and subsequent discovery, in 2015, of superconductivity at $T_c =200$~K in H$_3$S at one $\sim \SI{150}{GPa}$ pressure~\cite{Drozdov_H3S_2015}. This experiment shattered the assumed ceiling for $T_c$ in electron-phonon superconductors and was followed by the recent discovery of superconductivity in compressed lanthanum hydride at 215-250~K~\cite{Drozdov2018, somayazulu2019evidence}, also preceded by a theoretical prediction~\cite{liu_potential_2017,Peng2017}.  Recent computational approaches to hydride superconductivity have been reviewed in Refs.~\onlinecite{Boeri2019,PickettEremets2019}.

Despite these undeniable successes and the  demonstration that the old assumed limit of 35-40 K for $T_c$ due to the exchange of phonons is incorrect,    these experiments do not  provide a clear strategy to optimize $T_c$ in the vast phase space of materials. This is at least partially due to an inability to  identify the correct materials descriptors, parameters directly reflecting the underlying mechanism of superconductivity.  For some classes of materials, {\it e.g.}, thermoelectrics, considerable progress has been made in high-throughput approaches identifying simple observables recorded in databases that contribute to a material's figure of merit~\cite{curtarolo2013high}. For superconductivity, however, such approaches\cite{Stanev2018} are considerably more difficult, both because the theory is more complex, and the figure of merit, $T_c$, depends extremely sensitively on the underlying interactions.  

This last difficulty is clear already from the Bardeen-Cooper-Schrieffer (BCS) theory of superconductivity~\cite{BCS_Theory}, among whose great successes was the proof that for weak attractive interactions, fermions pair with an instability that corresponds to an essential singularity in the dimensionless coupling constant $\lambda$, leading to the well-known expression,
\begin{equation}
    \label{eq:BCS}
    T_c\simeq 1.14 \, \omega_D\, e^{-\frac{1}{\lambda}},
\end{equation} 
where $\omega_D$ is the Debye frequency. BCS theory is successful because it predicts superconducting properties accurately in terms of measured $T_c$'s, but the essential singularity alone suggests that accurate calculations will be difficult. Besides, Eq.~\eqref{eq:BCS} is strictly valid only in the weak coupling limit $\lambda \ll 1$ and if the Coulomb interaction is neglected.

The inadequacy of the BCS expression for $T_c$ was already clear by the late 1960's, when McMillan~\cite{Mcmillan1968} introduced an improved formula based on Eliashberg theory~\cite{EliashbergInteractionBetweenElAndLatticeVibrInASC1960}, relating $T_c$ to a small number of physical quantities calculated from the effective electron-phonon interaction $\alpha^2F(\omega)$ that could in principle be extracted from tunneling data~\cite{McmillanRowell1965},
\begin{equation}
  T_c  \simeq \frac{\omega_D}{1.45}\exp\left(-\frac{1.04(1+\lambda)}{\lambda-\mu^\ast(1+0.62\lambda)} \right),\label{eq:McMillan}
\end{equation}
where $\mu^\ast$ is the Coulomb pseudopotential. This approximate formula led McMillan to predict a maximum $T_c$ in a given class of materials. Dynes~\cite{Dynes1972} later replaced the prefactor $\omega_D$/1.45 of the McMillan equation \eqref{eq:McMillan} with $\langle \omega \rangle/1.20$, where $\langle \omega \rangle$ is the first moment of the distribution $g(\omega) = 2/(\lambda \omega) \alpha^2F(\omega)$. 

Based on a reanalysis of Eliashberg theory and newly available computational checks in special cases, Allen and Dynes~\cite{Allen-Dynes1975} proposed an alternate approximate formula, 
\begin{equation}
T_c={\frac{f_1f_2 \omega_{\log}}{1.20}} \exp\left(-\frac{1.04(1+\lambda)}{\lambda-\mu^\ast(1+0.62\lambda)}\right), \label{eq:AD}
\end{equation}
where $f_1$ and $f_2$ are correction factors that depend on $\lambda,\mu^\ast,\omega_{\log}$, and $\bar \omega_2$, where $\bar \omega_2$ is the second moment of the phonon density of states $F(\omega)$. They showed that the expression \eqref{eq:AD} fit the $T_c$ of  a variety of superconductors known at the time, using data derived from tunneling, and that it implied the absence of any maximum $T_c$, except that caused by the competition between $\lambda$ and $\omega_{\log}\equiv \exp \langle \ln \omega\rangle$, where the average is taken over $g(\omega)$. Unlike the McMillan expression, which saturates to a constant value as $\lambda\rightarrow \infty$, the Allen-Dynes  equation obeys an asymptotic result of Eliashberg theory, that $T_c\sim \sqrt{\lambda}$ as $\lambda\rightarrow \infty$ with other parameters fixed.

The Allen-Dynes equation has played a crucial role in the discussion of high-temperature superconductivity and indeed is often used to extract quoted values of $\lambda$ in the literature for materials where tunneling data is not available. Nevertheless, it is important to recall that it has been derived from Eliashberg theory, which itself is implemented with various approximations, {\it e.g.}, the momentum dependence of the electron-phonon interaction was often  neglected in early studies. The full evaluation of the Eliashberg equation is computationally expensive and not currently suitable for high-throughput superconductor discovery. It would be highly desirable to develop an expression for $T_c$ that generalizes the Allen-Dynes equation and is applicable over an extensive range of parameters to guide such searches.

In this letter, we use modern machine learning techniques to critically examine the  Allen-Dynes equation in the context of  recently discovered materials.  These techniques are {\it analytical} in nature, meaning they search for analytical relations between a minimal set of features, {\it i.e.}, physical parameters, and the desired properties. Specifically, we apply the Sure-Independence Screening and Sparsifying Operator (SISSO) method~\cite{SISSO2018} to estimate  $T_c$ from $\lambda$, $\mu^\ast$, and $\omega_{\log}$ with the goal to obtain an equation of similar or enhanced performance to the one proposed by Allen and Dynes~\cite{Allen-Dynes1975}. 
We find that  we can improve on the Allen-Dynes fit to strong-coupling superconductors, with a smaller set of descriptors.  More interestingly, the approach identifies outliers like MgB$_2$, $T_c$=39 K, which suggests the importance of new physics essential to high $T_c$ that needs to be incorporated in an improved formula to guide the search for new electron-phonon superconductors in materials space.



\section{Methods}
\begin{figure}[t]
\includegraphics[width=\columnwidth]{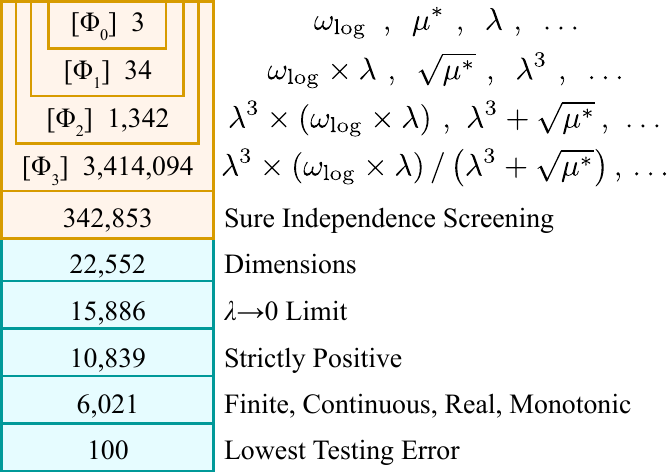}
\caption{Beginning with feature space $\Phi_0$, consisting of $\omega_{\log}$, $\lambda$ and $\mu^{\ast}$, each additional tier $\Phi_i$ is constructed by applying 4 binary operators (+, $-$, $\times$, /) and 7 unary operators (exp, log, $\;\sqrt{}$, $\;\sqrt[3]{}$, $\;^{-1}$, $\;^2$, $\;^3$) to features from preceding tiers. This procedure is applied up to level $\Phi_3$, after which sure-independence screening is applied to eliminate features with correlation factors (inner product) below 0.5 with respect to $T_c$. Physical constraints as listed are then applied to further reduce the feature space. We fit coefficients to the 6,021 features and obtain the 100 models with lowest root-mean-square error in predictions on the testing set.}
\label{fig:filtering}
\end{figure}

To generate models for predicting $T_c$, we apply recently developed methods of equation-based machine learning, subject to physical constraints.  In the SISSO approach, the predictive models are expressed as analytical formulas relating physical quantities  with algebraic operations such as addition and exponentiation. Given a tabulated set of scalar-valued physical quantities, or features, the SISSO method constructs additional features by iteratively applying operations from a specified set, {\it e.g.}, +, $\times$, exp, $\sqrt{~}$, $^2$.

\begin{figure*}[ht]
\centering
\includegraphics[width=\textwidth]{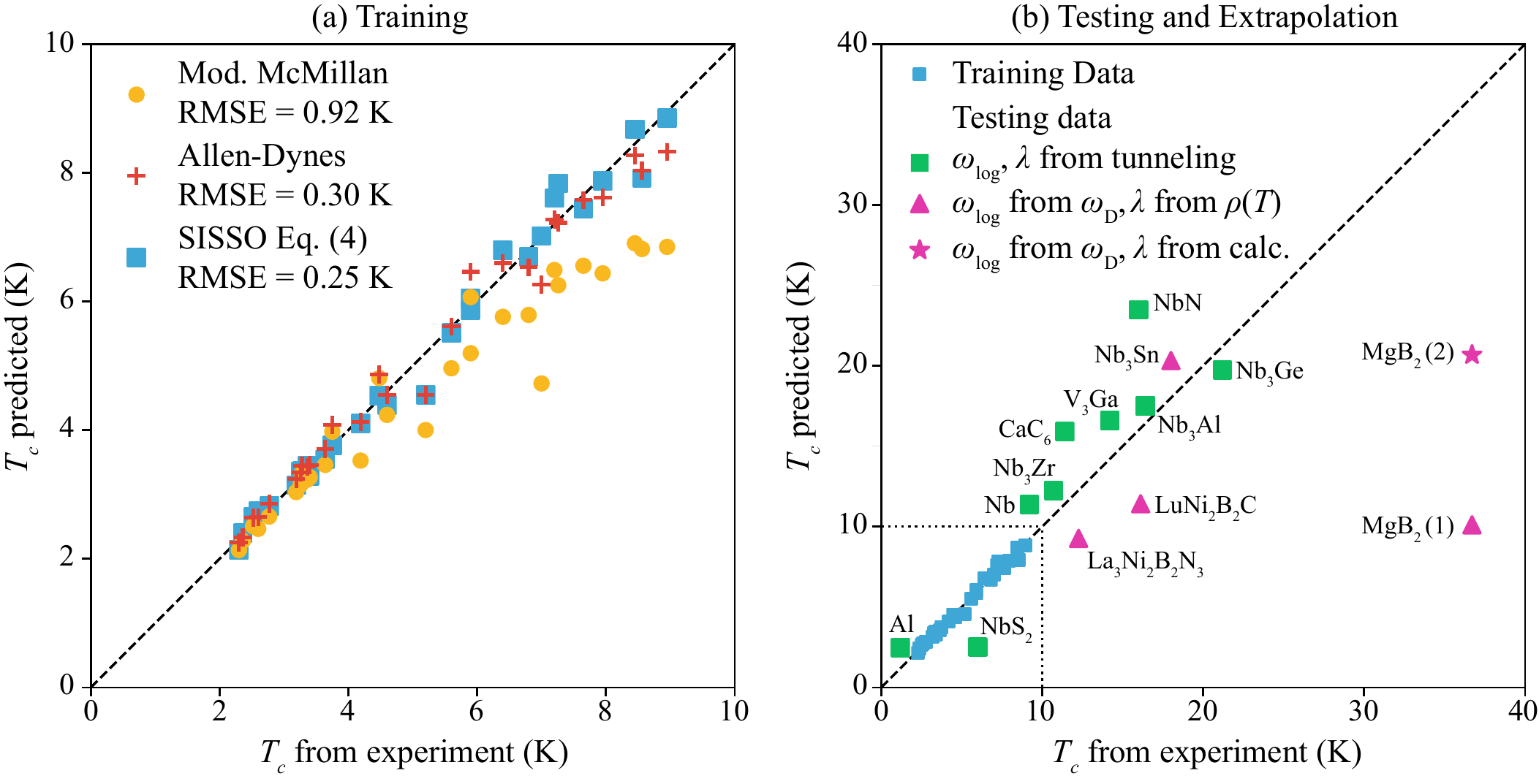}
\caption{Machine learning of optimal {\it analytical} expression for $T_c$ as a function of three parameters ($\omega_{\log}$, $\lambda$, and $\mu^{\ast}$) trained on the low-$T_c$ dataset of Allen and Dynes~\cite{Allen-Dynes1975} using the SISSO algorithm~\cite{SISSO2018}. (a) The 3-parameter machine-learned equation results in a smaller RMSE than the 4-parameter Allen-Dynes or the 3-parameter McMillan equation
(b) The testing of the machine-learned equation using 12 different superconductors assumes that $\mu^{\ast} = 0.1$ and takes $\omega_{\log}$ and $\lambda$ from tunneling measurements~\cite{Arnold1980, Mitrovic1984, Kim2006, Kwo1981, Bending1987, Kihlstrom1984, Nishio1993, Wolf1979, Kihlstrom1985}, except for Nb$_3$Sn, MgB$_2$(1), and the two quaternary compounds, for which $\omega_{\log}$ is obtained from low-temperature specific heat measurements and $\lambda$ from high-temperature resistivity~\cite{junod2001specific}, and for MgB$_2$(2) for which $\lambda$ is from density-functional calculations~\cite{Margine2013}. This extrapolation shows larger deviations with an  RMSE = 3.2 K or 9.1\% and two outliers, NbS$_2$ at low temperatures and MgB$_2$ at high temperatures.}
\label{fig:ML-AD}
\end{figure*}

To pinpoint the best equations, the SISSO method employs the sure-independence screening (SIS) method and the sparse-solution algorithm using sparsifying operators (SO) in tandem. After constructing the feature space, the SIS method selects a subspace of features with the largest linear correlation with the target property ($T_c$), {\it i.e.}, the largest absolute value of their dot product. The SO step then evaluates all possible combinations of features from the SIS subspace, yielding the optimal least-squares solution and residual. With such a vast feature space, the combinatorial optimization in each SO step relies on $L_0$ regularization, which penalizes the number of non-zero coefficients. Combined with one numerical prefactor, fit from available data, each feature is used to generate one predictive model. 

We benchmark the performance of different models identified by SISSO using leave-one-out cross-validation. Given $N$ available data points, each model is repeatedly fit using $N-1$ points and evaluated with the excluded point. The average evaluation error across $N$ iterations, where each point is tested once, is the leave-one-out cross-validation error. This method can help to maximize the transferability of a model by reducing ``overfitting", 
{\it i.e.}, models that exhibit low root-mean-square error in predictions on the training data but very high root-mean-square error in the testing data. 

We apply the SISSO method to estimate $T_c$ from $\lambda$, $\mu^\ast$, and $\omega_{\log}$ to obtain an equation of similar performance to the one proposed by Allen and Dynes~\cite{Allen-Dynes1975}. We use the values of $\lambda$, $\mu^\ast$, and $\omega_{\log}$, and the target property, $T_c$, from the data for 29 superconducting materials provided by Allen and Dynes (Table I in Ref.~\cite{Allen-Dynes1975}).   Next, we apply the SISSO method with 4 binary operators (+, $-$, $\times$, /) and 7 unary operators (exp, log, $\sqrt{}$, $\sqrt[3]{}$, $^{-1}$, $^2$, $^3$) three times to generate 3,414,094 features.  Fig.~\ref{fig:filtering} shows the rapid growth of the feature space with the number of iterations. Of the initial feature space, we select the equations with the highest linear correlation to $T_c$ using sure-independence screening with a minimum correlation magnitude (inner product) of 0.5. To further reduce the number of features and eliminate unphysical equations, we apply constraints. We select equations that are linearly proportional to $\omega_{\log}$ and obey the proper $\lambda\rightarrow 0$ limiting behavior. Additionally, we filter for equations that are strictly positive, real, finite, continuous, and monotonic across the relevant training and testing feature spaces. To evaluate the generalizability and performance of these equations, we compute the error against a testing set of 9 superconductors~\cite{Arnold1980, Mitrovic1984, Kim2006, Kwo1981, Bending1987, Kihlstrom1984, Nishio1993, Wolf1979, Kihlstrom1985}, shown in green in Fig. 2, using leave-one-out cross validation.

\section{Results}

Fig.~\ref{fig:ML-AD} illustrates the main proof-of-principle result that machine learning can provide an analytic equation of similar performance to the Allen-Dynes equation. The equation-based machine learning uses the values of $\lambda,\omega_{\log}$, and $\mu^\ast$ of the 29 materials in Table I of Allen and Dynes~\cite{Allen-Dynes1975}, and neglects the average frequency $\bar\omega_2\equiv \langle \omega^2 \rangle ^{1/2}$ taken over the $g(\omega)$ distribution that is also used in the Allen-Dynes equation.  The SISSO method and subsequent physical constraints lead to the optimal equation,
\begin{equation}
    {T_c}^{\mathrm{SISSO}} = 0.09525 \frac{\lambda^4  \omega_{\log}}{\lambda^3 + \sqrt{\mu^\ast}}.\label{eq:winner}
\end{equation}
Importantly, Eq.~\eqref{eq:winner} emerged from our approach with the smallest RMSE even before any of the physical constraints summarized in Fig.~\ref{fig:filtering} were  applied. Fig.~\ref{fig:ML-AD}(a) compares the performance of this equation with the modified McMillan and Allen-Dynes equations for the measured $T_c$'s of the 29 materials that train the model. The root-mean-square error (RMSE) of this equation evaluated on the training data is 0.25~K, significantly smaller than the RMSE of 0.92~K for the modified McMillan equation, and also slightly lower than the RMSE of 0.30~K for the Allen-Dynes equation. This result is impressive given the use of only 3 parameters and a single numerical coefficient compared to 3 parameters and 4 coefficients for the modified McMillan and 4 parameters and 7 coefficients for the Allen-Dynes equation.

Figure~\ref{fig:ML-AD}(b) shows the testing of Eq.~\eqref{eq:winner} for a variety of other superconductors, mostly of higher $T_c$. Because $\mu^\ast$ data were not available for these materials, we adopt a constant value of $\mu^\ast=0.1$. This procedure introduces some unknown error into the analysis, but despite this, the fit to the new materials is rather good, with an RMSE of only 3.2~K (9.1\%).

It is important to note that Eq.~\eqref{eq:winner} is not derived from any physical theory and therefore may contain some terms that may make no physical sense, {\it e.g.}, the appearance of the $\sqrt{\mu^\ast}$ term, which may be a proxy for a constant term due to the small range of data and the paucity of features at this level of learning. The limit $T_c\rightarrow 0$ as $\lambda\rightarrow 0$ in  Eq.~\eqref{eq:winner} even at nonzero $\mu^*$ may reflect the lack of data at small coupling. Also,  Eq.~\eqref{eq:winner} increases monotonically with $\lambda$, with linear behavior at very high couplings.  This behavior violates the asymptotic limit of Eliashberg theory, $T_c\sim \sqrt{\lambda}$, built into the Allen-Dynes equation~\cite{Allen-Dynes1975}. Again, this disagreement with physics is due to the absence of data points, either in the training or the testing set, which deviate significantly from the linear behavior predicted by Eq.~\eqref{eq:winner}.

\begin{figure}[t]
\centering\includegraphics[width=\columnwidth]{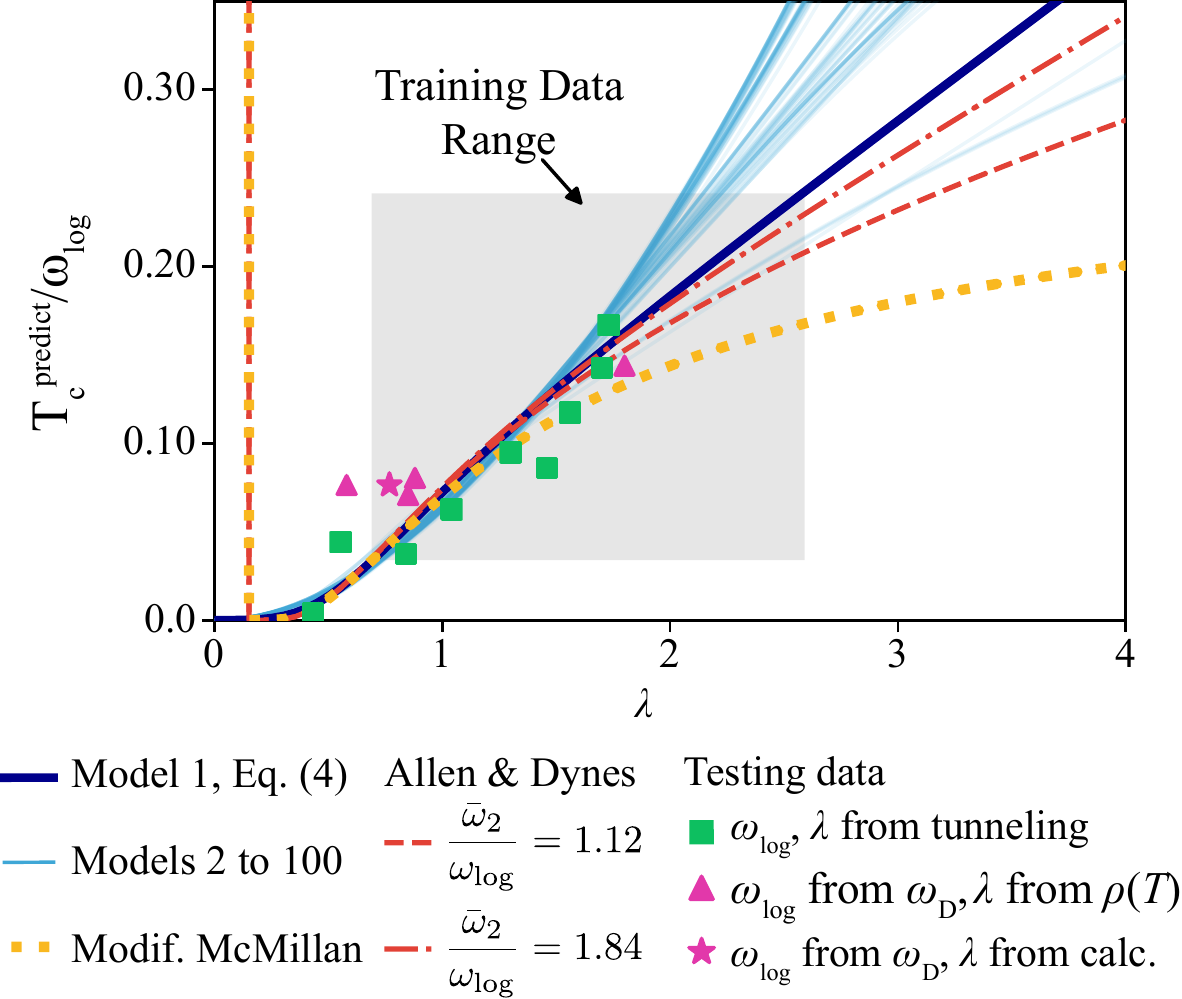}
\caption{$\lambda$ dependence of $T_c$ in the top 100 models, ranked by testing error assuming $\mu^\ast=0.1$. Two red curves correspond to the Allen-Dynes equation with the minimum and maximum values of $\bar\omega_2/\omega_{\log}$ in the training set. The modified McMillan equation systematically predicts smaller $T_c$'s and over the range of available $\lambda$ values, the simple machine-learned model closely matches the more complex Allen-Dynes equation.}
\label{fig:functional_form_lambda}
\end{figure}

Fig.~\ref{fig:functional_form_lambda} shows the functional behavior $T_c(\lambda)$ of the 100 highest-scored equations discovered by SISSO; it is clear that almost all of these equations are equally valid over the range of $\lambda$ values where data exist. This highlights the need for measurements to determine the materials parameters $\lambda$, $\omega_{\log}$, and $\mu^\ast$ reliably for both very low $T_c$ materials, as well as for some of the recently discovered higher-$T_c$ systems.

Fig.~\ref{fig:ML-AD}(b) also shows some dramatic failures of the learned equation, namely for MgB$_2$ and NbS$_2$. The probable reasons for these failures are both revealing and reassuring.  The point labeled MgB$_2$(1) with a predicted $T_c$ of 10~K is one where $\omega_{\log}$, a logarithmic average of the electron-phonon interaction function $\alpha^2F/\omega$, was determined from a specific heat measurement of the Debye frequency $\omega_D$, which depends only on the phonon density of states $F(\omega)$. Relating the Debye frequency with $\omega_{\log}$ neglects the difference between the two distributions~\cite{Allen-Dynes1975}.
This assumption is particularly poor in MgB$_2$, where high-frequency phonons couple anomalously strongly. In addition, $\lambda$ was determined from standard expressions for the high-temperature resistivity of a 3D metal.  It is well known that MgB$_2$ has strong 2D character, and that the full momentum and band dependence of the Eliashberg function $\lambda_{n{\bf k},n{\bf k'}}$ must be accounted for to obtain reasonable values for $T_c$ from first principles~\cite{Margine2013}. 
It is interesting to note that if one uses the higher value of $\lambda$ obtained from Ref.~\onlinecite{Margine2013} in Eq.~\eqref{eq:winner}, one obtains data point MgB$_2$(2), with the significantly enhanced predicted $T_c$ of 20~K, but still far from the measured value of 40~K and even further from the full Eliashberg calculation of 50~K~\cite{Margine2013}.

These discrepancies indicate, not surprisingly, that a machine trained on a database of nearly isotropic low-$T_c$ superconductors cannot capture the physics of highly anisotropic higher-$T_c$ materials using the simple averaged descriptors chosen by Allen and Dynes.  The same principle apparently applies to NbS$_2$, which while having a low-$T_c$ is quite 2-dimensional. However, Eq.~\eqref{eq:winner} may have significant predictive power extrapolated to higher-$T_c$ 3D systems.  For example, if we take values of $\lambda$ and $\omega_{\log}$ calculated from first principles for H$_3$S at  140~GPa pressure \cite{Duan14,Krakauer2015}, the predicted $T_c$  from Eq. (\ref{eq:winner}) is 262~K, compared to the measured value of 203~K.  This result is similar to the result obtained from Allen-Dynes using calculated parameters, but substantially higher than the modified McMillan equation used in Refs.~\onlinecite{Duan14,Krakauer2015}.

\section{Conclusion}
We have demonstrated that machine learning can discover equations that describe the dependence of superconducting $T_c$'s on moments of distributions of phonon frequencies and electron-phonon couplings,  as used originally by Allen and Dynes in their attempt to understand the systematics of $T_c$ in the framework of Eliashberg theory.  While the method is quite successful in predicting known  superconductors of the same general type as the original Allen-Dynes dataset, with fewer parameters, the existence of a few anomalous outliers suggests that the use of such methods for high-throughput materials discovery will require new descriptors that capture anomalous features, {\it e.g.}, the anisotropy of the electron-phonon interactions and unusual electronic states that take advantage of them.  A natural modern extension of the philosophy of Allen and Dynes is then to calculate from first principles a few key measures of electronic structure crucial for superconductivity, together with the moments discussed above, and apply machine-learning methods as described here.  We anticipate that this approach will allow a much more efficient and thorough investigation of materials space than current approaches that rely on fully anisotropic Eliashberg calculations for each material.  


\section{Acknowledgements}
We are grateful to L. Boeri, P. Allen, and W. Pickett for valuable discussions.  The work by SRX and RGH was supported by the National Science Foundation under grants Nos.\ PHY-1549132 and DMR-1609306. The work by GRS 
was performed under the auspices of the Office of Sciences, United States Department of Energy, contract numbers DE-FG02-86ER45268. 

\bibliographystyle{apsrev4-1}
\bibliography{references}

\begin{thebibliography}{32}%
\makeatletter
\providecommand \@ifxundefined [1]{%
 \@ifx{#1\undefined}
}%
\providecommand \@ifnum [1]{%
 \ifnum #1\expandafter \@firstoftwo
 \else \expandafter \@secondoftwo
 \fi
}%
\providecommand \@ifx [1]{%
 \ifx #1\expandafter \@firstoftwo
 \else \expandafter \@secondoftwo
 \fi
}%
\providecommand \natexlab [1]{#1}%
\providecommand \enquote  [1]{``#1''}%
\providecommand \bibnamefont  [1]{#1}%
\providecommand \bibfnamefont [1]{#1}%
\providecommand \citenamefont [1]{#1}%
\providecommand \href@noop [0]{\@secondoftwo}%
\providecommand \href [0]{\begingroup \@sanitize@url \@href}%
\providecommand \@href[1]{\@@startlink{#1}\@@href}%
\providecommand \@@href[1]{\endgroup#1\@@endlink}%
\providecommand \@sanitize@url [0]{\catcode `\\12\catcode `\$12\catcode
  `\&12\catcode `\#12\catcode `\^12\catcode `\_12\catcode `\%12\relax}%
\providecommand \@@startlink[1]{}%
\providecommand \@@endlink[0]{}%
\providecommand \url  [0]{\begingroup\@sanitize@url \@url }%
\providecommand \@url [1]{\endgroup\@href {#1}{\urlprefix }}%
\providecommand \urlprefix  [0]{URL }%
\providecommand \Eprint [0]{\href }%
\providecommand \doibase [0]{http://dx.doi.org/}%
\providecommand \selectlanguage [0]{\@gobble}%
\providecommand \bibinfo  [0]{\@secondoftwo}%
\providecommand \bibfield  [0]{\@secondoftwo}%
\providecommand \translation [1]{[#1]}%
\providecommand \BibitemOpen [0]{}%
\providecommand \bibitemStop [0]{}%
\providecommand \bibitemNoStop [0]{.\EOS\space}%
\providecommand \EOS [0]{\spacefactor3000\relax}%
\providecommand \BibitemShut  [1]{\csname bibitem#1\endcsname}%
\let\auto@bib@innerbib\@empty
\bibitem [{\citenamefont {Norman}(2016)}]{genome_SC}%
  \BibitemOpen
  \bibfield  {author} {\bibinfo {author} {\bibfnamefont {M.~R.}\ \bibnamefont
  {Norman}},\ }\href@noop {} {\bibfield  {journal} {\bibinfo  {journal} {Rep.
  Prog. Phys.}\ }\textbf {\bibinfo {volume} {79}},\ \bibinfo {pages} {074502}
  (\bibinfo {year} {2016})}\BibitemShut {NoStop}%
\bibitem [{\citenamefont {Pickett}(2017)}]{Pickett2018}%
  \BibitemOpen
  \bibfield  {author} {\bibinfo {author} {\bibfnamefont {W.~E.}\ \bibnamefont
  {Pickett}},\ }\href@noop {} {\bibfield  {journal} {\bibinfo  {journal} {arXiv
  preprint arXiv:1801.00165}\ } (\bibinfo {year} {2017})}\BibitemShut {NoStop}%
\bibitem [{\citenamefont {Duan}\ \emph {et~al.}(2019)\citenamefont {Duan},
  \citenamefont {Yu}, \citenamefont {Xie},\ and\ \citenamefont
  {Cui}}]{Duan_review2019}%
  \BibitemOpen
  \bibfield  {author} {\bibinfo {author} {\bibfnamefont {D.}~\bibnamefont
  {Duan}}, \bibinfo {author} {\bibfnamefont {H.}~\bibnamefont {Yu}}, \bibinfo
  {author} {\bibfnamefont {H.}~\bibnamefont {Xie}}, \ and\ \bibinfo {author}
  {\bibfnamefont {T.}~\bibnamefont {Cui}},\ }\href {\doibase
  https://doi.org/10.1007/s10948-018-4900-8} {\bibfield  {journal} {\bibinfo
  {journal} {J. Supercond. Novel Mag.}\ }\textbf {\bibinfo {volume} {32}},\
  \bibinfo {pages} {53} (\bibinfo {year} {2019})}\BibitemShut {NoStop}%
\bibitem [{\citenamefont {Drozdov}\ \emph {et~al.}(2015)\citenamefont
  {Drozdov}, \citenamefont {Eremets}, \citenamefont {Troyan}, \citenamefont
  {Ksenofontov},\ and\ \citenamefont {Shylin}}]{Drozdov_H3S_2015}%
  \BibitemOpen
  \bibfield  {author} {\bibinfo {author} {\bibfnamefont {A.~P.}\ \bibnamefont
  {Drozdov}}, \bibinfo {author} {\bibfnamefont {M.~I.}\ \bibnamefont
  {Eremets}}, \bibinfo {author} {\bibfnamefont {I.~A.}\ \bibnamefont {Troyan}},
  \bibinfo {author} {\bibfnamefont {V.}~\bibnamefont {Ksenofontov}}, \ and\
  \bibinfo {author} {\bibfnamefont {S.~I.}\ \bibnamefont {Shylin}},\ }\href
  {\doibase 10.1038/nature14964} {\bibfield  {journal} {\bibinfo  {journal}
  {Nature}\ }\textbf {\bibinfo {volume} {525}},\ \bibinfo {pages} {73}
  (\bibinfo {year} {2015})}\BibitemShut {NoStop}%
\bibitem [{\citenamefont {Drozdov}\ \emph {et~al.}(2018)\citenamefont
  {Drozdov}, \citenamefont {Minkov}, \citenamefont {Besedin}, \citenamefont
  {Kong}, \citenamefont {Kuzovnikov}, \citenamefont {Knyazev},\ and\
  \citenamefont {Eremets}}]{Drozdov2018}%
  \BibitemOpen
  \bibfield  {author} {\bibinfo {author} {\bibfnamefont {A.}~\bibnamefont
  {Drozdov}}, \bibinfo {author} {\bibfnamefont {V.}~\bibnamefont {Minkov}},
  \bibinfo {author} {\bibfnamefont {S.}~\bibnamefont {Besedin}}, \bibinfo
  {author} {\bibfnamefont {P.}~\bibnamefont {Kong}}, \bibinfo {author}
  {\bibfnamefont {M.}~\bibnamefont {Kuzovnikov}}, \bibinfo {author}
  {\bibfnamefont {D.}~\bibnamefont {Knyazev}}, \ and\ \bibinfo {author}
  {\bibfnamefont {M.}~\bibnamefont {Eremets}},\ }\href@noop {} {\bibfield
  {journal} {\bibinfo  {journal} {arXiv preprint arXiv:1808.07039}\ } (\bibinfo
  {year} {2018})}\BibitemShut {NoStop}%
\bibitem [{\citenamefont {Somayazulu}\ \emph {et~al.}(2019)\citenamefont
  {Somayazulu}, \citenamefont {Ahart}, \citenamefont {Mishra}, \citenamefont
  {Geballe}, \citenamefont {Baldini}, \citenamefont {Meng}, \citenamefont
  {Struzhkin},\ and\ \citenamefont {Hemley}}]{somayazulu2019evidence}%
  \BibitemOpen
  \bibfield  {author} {\bibinfo {author} {\bibfnamefont {M.}~\bibnamefont
  {Somayazulu}}, \bibinfo {author} {\bibfnamefont {M.}~\bibnamefont {Ahart}},
  \bibinfo {author} {\bibfnamefont {A.~K.}\ \bibnamefont {Mishra}}, \bibinfo
  {author} {\bibfnamefont {Z.~M.}\ \bibnamefont {Geballe}}, \bibinfo {author}
  {\bibfnamefont {M.}~\bibnamefont {Baldini}}, \bibinfo {author} {\bibfnamefont
  {Y.}~\bibnamefont {Meng}}, \bibinfo {author} {\bibfnamefont {V.~V.}\
  \bibnamefont {Struzhkin}}, \ and\ \bibinfo {author} {\bibfnamefont {R.~J.}\
  \bibnamefont {Hemley}},\ }\href {\doibase 10.1103/PhysRevLett.122.027001}
  {\bibfield  {journal} {\bibinfo  {journal} {Phys. Rev. Lett.}\ }\textbf
  {\bibinfo {volume} {122}},\ \bibinfo {pages} {027001} (\bibinfo {year}
  {2019})}\BibitemShut {NoStop}%
\bibitem [{\citenamefont {Liu}\ \emph {et~al.}(2017)\citenamefont {Liu},
  \citenamefont {Naumov}, \citenamefont {Hoffmann}, \citenamefont {Ashcroft},\
  and\ \citenamefont {Hemley}}]{liu_potential_2017}%
  \BibitemOpen
  \bibfield  {author} {\bibinfo {author} {\bibfnamefont {H.}~\bibnamefont
  {Liu}}, \bibinfo {author} {\bibfnamefont {I.~I.}\ \bibnamefont {Naumov}},
  \bibinfo {author} {\bibfnamefont {R.}~\bibnamefont {Hoffmann}}, \bibinfo
  {author} {\bibfnamefont {N.~W.}\ \bibnamefont {Ashcroft}}, \ and\ \bibinfo
  {author} {\bibfnamefont {R.~J.}\ \bibnamefont {Hemley}},\ }\href {\doibase
  10.1073/pnas.1704505114} {\bibfield  {journal} {\bibinfo  {journal} {Proc.
  Natl. Acad. Sci. U.S.A.}\ }\textbf {\bibinfo {volume} {114}},\ \bibinfo
  {pages} {6990} (\bibinfo {year} {2017})}\BibitemShut {NoStop}%
\bibitem [{\citenamefont {Peng}\ \emph {et~al.}(2017)\citenamefont {Peng},
  \citenamefont {Sun}, \citenamefont {Pickard}, \citenamefont {Needs},
  \citenamefont {Wu},\ and\ \citenamefont {Ma}}]{Peng2017}%
  \BibitemOpen
  \bibfield  {author} {\bibinfo {author} {\bibfnamefont {F.}~\bibnamefont
  {Peng}}, \bibinfo {author} {\bibfnamefont {Y.}~\bibnamefont {Sun}}, \bibinfo
  {author} {\bibfnamefont {C.~J.}\ \bibnamefont {Pickard}}, \bibinfo {author}
  {\bibfnamefont {R.~J.}\ \bibnamefont {Needs}}, \bibinfo {author}
  {\bibfnamefont {Q.}~\bibnamefont {Wu}}, \ and\ \bibinfo {author}
  {\bibfnamefont {Y.}~\bibnamefont {Ma}},\ }\href {\doibase
  10.1103/PhysRevLett.119.107001} {\bibfield  {journal} {\bibinfo  {journal}
  {Phys. Rev. Lett.}\ }\textbf {\bibinfo {volume} {119}},\ \bibinfo {pages}
  {107001} (\bibinfo {year} {2017})}\BibitemShut {NoStop}%
\bibitem [{\citenamefont {{Boeri}}(2019)}]{Boeri2019}%
  \BibitemOpen
  \bibfield  {author} {\bibinfo {author} {\bibfnamefont {L.}~\bibnamefont
  {{Boeri}}},\ }\href@noop {} {\bibfield  {journal} {\bibinfo  {journal} {arXiv
  e-prints}\ ,\ \bibinfo {eid} {arXiv:1903.05708}} (\bibinfo {year} {2019})},\
  \Eprint {http://arxiv.org/abs/1903.05708} {arXiv:1903.05708
  [cond-mat.supr-con]} \BibitemShut {NoStop}%
\bibitem [{\citenamefont {{Pickett}}\ and\ \citenamefont
  {{Eremets}}(2019)}]{PickettEremets2019}%
  \BibitemOpen
  \bibfield  {author} {\bibinfo {author} {\bibfnamefont {W.}~\bibnamefont
  {{Pickett}}}\ and\ \bibinfo {author} {\bibfnamefont {M.}~\bibnamefont
  {{Eremets}}},\ }\href {\doibase 10.1063/PT.3.4204} {\bibfield  {journal}
  {\bibinfo  {journal} {Physics Today}\ }\textbf {\bibinfo {volume} {72}},\
  \bibinfo {pages} {52} (\bibinfo {year} {2019})}\BibitemShut {NoStop}%
\bibitem [{\citenamefont {Curtarolo}\ \emph {et~al.}(2013)\citenamefont
  {Curtarolo}, \citenamefont {Hart}, \citenamefont {Nardelli}, \citenamefont
  {Mingo}, \citenamefont {Sanvito},\ and\ \citenamefont
  {Levy}}]{curtarolo2013high}%
  \BibitemOpen
  \bibfield  {author} {\bibinfo {author} {\bibfnamefont {S.}~\bibnamefont
  {Curtarolo}}, \bibinfo {author} {\bibfnamefont {G.~L.}\ \bibnamefont {Hart}},
  \bibinfo {author} {\bibfnamefont {M.~B.}\ \bibnamefont {Nardelli}}, \bibinfo
  {author} {\bibfnamefont {N.}~\bibnamefont {Mingo}}, \bibinfo {author}
  {\bibfnamefont {S.}~\bibnamefont {Sanvito}}, \ and\ \bibinfo {author}
  {\bibfnamefont {O.}~\bibnamefont {Levy}},\ }\href@noop {} {\bibfield
  {journal} {\bibinfo  {journal} {Nat. Mater.}\ }\textbf {\bibinfo {volume}
  {12}},\ \bibinfo {pages} {191} (\bibinfo {year} {2013})}\BibitemShut
  {NoStop}%
\bibitem [{\citenamefont {{Stanev}}\ \emph {et~al.}(2018)\citenamefont
  {{Stanev}}, \citenamefont {{Oses}}, \citenamefont {{Kusne}}, \citenamefont
  {{Rodriguez}}, \citenamefont {{Paglione}}, \citenamefont {{Curtarolo}},\ and\
  \citenamefont {{Takeuchi}}}]{Stanev2018}%
  \BibitemOpen
  \bibfield  {author} {\bibinfo {author} {\bibfnamefont {V.}~\bibnamefont
  {{Stanev}}}, \bibinfo {author} {\bibfnamefont {C.}~\bibnamefont {{Oses}}},
  \bibinfo {author} {\bibfnamefont {A.~G.}\ \bibnamefont {{Kusne}}}, \bibinfo
  {author} {\bibfnamefont {E.}~\bibnamefont {{Rodriguez}}}, \bibinfo {author}
  {\bibfnamefont {J.}~\bibnamefont {{Paglione}}}, \bibinfo {author}
  {\bibfnamefont {S.}~\bibnamefont {{Curtarolo}}}, \ and\ \bibinfo {author}
  {\bibfnamefont {I.}~\bibnamefont {{Takeuchi}}},\ }\href {\doibase
  10.1038/s41524-018-0085-8} {\bibfield  {journal} {\bibinfo  {journal} {npj
  Comput. Mater.}\ }\textbf {\bibinfo {volume} {4}},\ \bibinfo {eid} {29}
  (\bibinfo {year} {2018})}\BibitemShut {NoStop}%
\bibitem [{\citenamefont {Bardeen}\ \emph {et~al.}(1957)\citenamefont
  {Bardeen}, \citenamefont {Cooper},\ and\ \citenamefont
  {Schrieffer}}]{BCS_Theory}%
  \BibitemOpen
  \bibfield  {author} {\bibinfo {author} {\bibfnamefont {J.}~\bibnamefont
  {Bardeen}}, \bibinfo {author} {\bibfnamefont {L.~N.}\ \bibnamefont {Cooper}},
  \ and\ \bibinfo {author} {\bibfnamefont {J.~R.}\ \bibnamefont {Schrieffer}},\
  }\href {\doibase 10.1103/PhysRev.108.1175} {\bibfield  {journal} {\bibinfo
  {journal} {Phys. Rev.}\ }\textbf {\bibinfo {volume} {108}},\ \bibinfo {pages}
  {1175} (\bibinfo {year} {1957})}\BibitemShut {NoStop}%
\bibitem [{\citenamefont {McMillan}(1968)}]{Mcmillan1968}%
  \BibitemOpen
  \bibfield  {author} {\bibinfo {author} {\bibfnamefont {W.~L.}\ \bibnamefont
  {McMillan}},\ }\href {\doibase 10.1103/PhysRev.167.331} {\bibfield  {journal}
  {\bibinfo  {journal} {Phys. Rev.}\ }\textbf {\bibinfo {volume} {167}},\
  \bibinfo {pages} {331} (\bibinfo {year} {1968})}\BibitemShut {NoStop}%
\bibitem [{\citenamefont
  {Eliashberg}(1960)}]{EliashbergInteractionBetweenElAndLatticeVibrInASC1960}%
  \BibitemOpen
  \bibfield  {author} {\bibinfo {author} {\bibfnamefont {G.~M.}\ \bibnamefont
  {Eliashberg}},\ }\href@noop {} {\bibfield  {journal} {\bibinfo  {journal}
  {Sov. Phys. JETP}\ }\textbf {\bibinfo {volume} {11}},\ \bibinfo {pages} {696}
  (\bibinfo {year} {1960})}\BibitemShut {NoStop}%
\bibitem [{\citenamefont {McMillan}\ and\ \citenamefont
  {Rowell}(1965)}]{McmillanRowell1965}%
  \BibitemOpen
  \bibfield  {author} {\bibinfo {author} {\bibfnamefont {W.~L.}\ \bibnamefont
  {McMillan}}\ and\ \bibinfo {author} {\bibfnamefont {J.~M.}\ \bibnamefont
  {Rowell}},\ }\href {\doibase 10.1103/PhysRevLett.14.108} {\bibfield
  {journal} {\bibinfo  {journal} {Phys. Rev. Lett.}\ }\textbf {\bibinfo
  {volume} {14}},\ \bibinfo {pages} {108} (\bibinfo {year} {1965})}\BibitemShut
  {NoStop}%
\bibitem [{\citenamefont {Dynes}(1972)}]{Dynes1972}%
  \BibitemOpen
  \bibfield  {author} {\bibinfo {author} {\bibfnamefont {R.}~\bibnamefont
  {Dynes}},\ }\href {\doibase https://doi.org/10.1016/0038-1098(72)90603-5}
  {\bibfield  {journal} {\bibinfo  {journal} {Solid State Commun.}\ }\textbf
  {\bibinfo {volume} {10}},\ \bibinfo {pages} {615 } (\bibinfo {year}
  {1972})}\BibitemShut {NoStop}%
\bibitem [{\citenamefont {Allen}\ and\ \citenamefont
  {Dynes}(1975)}]{Allen-Dynes1975}%
  \BibitemOpen
  \bibfield  {author} {\bibinfo {author} {\bibfnamefont {P.~B.}\ \bibnamefont
  {Allen}}\ and\ \bibinfo {author} {\bibfnamefont {R.~C.}\ \bibnamefont
  {Dynes}},\ }\href {\doibase 10.1103/PhysRevB.12.905} {\bibfield  {journal}
  {\bibinfo  {journal} {Phys. Rev. B}\ }\textbf {\bibinfo {volume} {12}},\
  \bibinfo {pages} {905} (\bibinfo {year} {1975})}\BibitemShut {NoStop}%
\bibitem [{\citenamefont {Ouyang}\ \emph {et~al.}(2018)\citenamefont {Ouyang},
  \citenamefont {Curtarolo}, \citenamefont {Ahmetcik}, \citenamefont
  {Scheffler},\ and\ \citenamefont {Ghiringhelli}}]{SISSO2018}%
  \BibitemOpen
  \bibfield  {author} {\bibinfo {author} {\bibfnamefont {R.}~\bibnamefont
  {Ouyang}}, \bibinfo {author} {\bibfnamefont {S.}~\bibnamefont {Curtarolo}},
  \bibinfo {author} {\bibfnamefont {E.}~\bibnamefont {Ahmetcik}}, \bibinfo
  {author} {\bibfnamefont {M.}~\bibnamefont {Scheffler}}, \ and\ \bibinfo
  {author} {\bibfnamefont {L.~M.}\ \bibnamefont {Ghiringhelli}},\ }\href
  {\doibase 10.1103/PhysRevMaterials.2.083802} {\bibfield  {journal} {\bibinfo
  {journal} {Phys. Rev. Materials}\ }\textbf {\bibinfo {volume} {2}},\ \bibinfo
  {pages} {083802} (\bibinfo {year} {2018})}\BibitemShut {NoStop}%
\bibitem [{\citenamefont {Arnold}\ \emph {et~al.}(1980)\citenamefont {Arnold},
  \citenamefont {Zasadzinski}, \citenamefont {Osmun},\ and\ \citenamefont
  {Wolf}}]{Arnold1980}%
  \BibitemOpen
  \bibfield  {author} {\bibinfo {author} {\bibfnamefont {G.~B.}\ \bibnamefont
  {Arnold}}, \bibinfo {author} {\bibfnamefont {J.}~\bibnamefont {Zasadzinski}},
  \bibinfo {author} {\bibfnamefont {J.~W.}\ \bibnamefont {Osmun}}, \ and\
  \bibinfo {author} {\bibfnamefont {E.~L.}\ \bibnamefont {Wolf}},\ }\href
  {\doibase 10.1007/BF00117117} {\bibfield  {journal} {\bibinfo  {journal} {J.
  Low Temp. Phys.}\ }\textbf {\bibinfo {volume} {40}},\ \bibinfo {pages} {225}
  (\bibinfo {year} {1980})}\BibitemShut {NoStop}%
\bibitem [{\citenamefont {Mitrovi{\'{c}}}\ \emph {et~al.}(1984)\citenamefont
  {Mitrovi{\'{c}}}, \citenamefont {Zarate},\ and\ \citenamefont
  {Carbotte}}]{Mitrovic1984}%
  \BibitemOpen
  \bibfield  {author} {\bibinfo {author} {\bibfnamefont {B.}~\bibnamefont
  {Mitrovi{\'{c}}}}, \bibinfo {author} {\bibfnamefont {H.~G.}\ \bibnamefont
  {Zarate}}, \ and\ \bibinfo {author} {\bibfnamefont {J.~P.}\ \bibnamefont
  {Carbotte}},\ }\href {\doibase 10.1103/PhysRevB.29.184} {\bibfield  {journal}
  {\bibinfo  {journal} {Phys. Rev. B}\ }\textbf {\bibinfo {volume} {29}},\
  \bibinfo {pages} {184} (\bibinfo {year} {1984})}\BibitemShut {NoStop}%
\bibitem [{\citenamefont {Kim}\ \emph {et~al.}(2006)\citenamefont {Kim},
  \citenamefont {Boeri}, \citenamefont {Kremer},\ and\ \citenamefont
  {Razavi}}]{Kim2006}%
  \BibitemOpen
  \bibfield  {author} {\bibinfo {author} {\bibfnamefont {J.~S.}\ \bibnamefont
  {Kim}}, \bibinfo {author} {\bibfnamefont {L.}~\bibnamefont {Boeri}}, \bibinfo
  {author} {\bibfnamefont {R.~K.}\ \bibnamefont {Kremer}}, \ and\ \bibinfo
  {author} {\bibfnamefont {F.~S.}\ \bibnamefont {Razavi}},\ }\href {\doibase
  10.1103/PhysRevB.74.214513} {\bibfield  {journal} {\bibinfo  {journal} {Phys.
  Rev. B}\ }\textbf {\bibinfo {volume} {74}},\ \bibinfo {pages} {1} (\bibinfo
  {year} {2006})}\BibitemShut {NoStop}%
\bibitem [{\citenamefont {Kwo}\ and\ \citenamefont {Geballe}(1981)}]{Kwo1981}%
  \BibitemOpen
  \bibfield  {author} {\bibinfo {author} {\bibfnamefont {J.}~\bibnamefont
  {Kwo}}\ and\ \bibinfo {author} {\bibfnamefont {T.~H.}\ \bibnamefont
  {Geballe}},\ }\href {\doibase 10.1103/PhysRevB.23.3230} {\bibfield  {journal}
  {\bibinfo  {journal} {Phys. Rev. B}\ }\textbf {\bibinfo {volume} {23}},\
  \bibinfo {pages} {3230} (\bibinfo {year} {1981})}\BibitemShut {NoStop}%
\bibitem [{\citenamefont {Bending}\ \emph {et~al.}(1987)\citenamefont
  {Bending}, \citenamefont {Beasley},\ and\ \citenamefont
  {Wolf}}]{Bending1987}%
  \BibitemOpen
  \bibfield  {author} {\bibinfo {author} {\bibfnamefont {S.~J.}\ \bibnamefont
  {Bending}}, \bibinfo {author} {\bibfnamefont {M.~R.}\ \bibnamefont
  {Beasley}}, \ and\ \bibinfo {author} {\bibfnamefont {E.~L.}\ \bibnamefont
  {Wolf}},\ }\href {\doibase 10.1103/PhysRevB.35.115} {\bibfield  {journal}
  {\bibinfo  {journal} {Phys. Rev. B}\ }\textbf {\bibinfo {volume} {35}},\
  \bibinfo {pages} {115} (\bibinfo {year} {1987})}\BibitemShut {NoStop}%
\bibitem [{\citenamefont {Kihlstrom}\ \emph {et~al.}(1984)\citenamefont
  {Kihlstrom}, \citenamefont {Mael},\ and\ \citenamefont
  {Geballe}}]{Kihlstrom1984}%
  \BibitemOpen
  \bibfield  {author} {\bibinfo {author} {\bibfnamefont {K.~E.}\ \bibnamefont
  {Kihlstrom}}, \bibinfo {author} {\bibfnamefont {D.}~\bibnamefont {Mael}}, \
  and\ \bibinfo {author} {\bibfnamefont {T.~H.}\ \bibnamefont {Geballe}},\
  }\href {\doibase 10.1103/PhysRevB.29.150} {\bibfield  {journal} {\bibinfo
  {journal} {Phys. Rev. B}\ }\textbf {\bibinfo {volume} {29}} (\bibinfo {year}
  {1984}),\ 10.1103/PhysRevB.29.150}\BibitemShut {NoStop}%
\bibitem [{\citenamefont {Nishio}\ \emph {et~al.}(1993)\citenamefont {Nishio},
  \citenamefont {Shirai}, \citenamefont {Suzuki},\ and\ \citenamefont
  {Motizuki}}]{Nishio1993}%
  \BibitemOpen
  \bibfield  {author} {\bibinfo {author} {\bibfnamefont {Y.}~\bibnamefont
  {Nishio}}, \bibinfo {author} {\bibfnamefont {M.}~\bibnamefont {Shirai}},
  \bibinfo {author} {\bibfnamefont {N.}~\bibnamefont {Suzuki}}, \ and\ \bibinfo
  {author} {\bibfnamefont {K.}~\bibnamefont {Motizuki}},\ }\href@noop {}
  {\bibfield  {journal} {\bibinfo  {journal} {Int. J. Mod. Phys. B}\ }\textbf
  {\bibinfo {volume} {7}},\ \bibinfo {pages} {188} (\bibinfo {year}
  {1993})}\BibitemShut {NoStop}%
\bibitem [{\citenamefont {Wolf}\ and\ \citenamefont {Noer}(1979)}]{Wolf1979}%
  \BibitemOpen
  \bibfield  {author} {\bibinfo {author} {\bibfnamefont {E.~L.}\ \bibnamefont
  {Wolf}}\ and\ \bibinfo {author} {\bibfnamefont {R.~J.}\ \bibnamefont
  {Noer}},\ }\href {\doibase 10.1016/0038-1098(79)90658-6} {\bibfield
  {journal} {\bibinfo  {journal} {Solid State Commun.}\ }\textbf {\bibinfo
  {volume} {30}},\ \bibinfo {pages} {391} (\bibinfo {year} {1979})}\BibitemShut
  {NoStop}%
\bibitem [{\citenamefont {Kihlstrom}\ \emph {et~al.}(1985)\citenamefont
  {Kihlstrom}, \citenamefont {Simon},\ and\ \citenamefont
  {Wolf}}]{Kihlstrom1985}%
  \BibitemOpen
  \bibfield  {author} {\bibinfo {author} {\bibfnamefont {K.~E.}\ \bibnamefont
  {Kihlstrom}}, \bibinfo {author} {\bibfnamefont {R.~W.}\ \bibnamefont
  {Simon}}, \ and\ \bibinfo {author} {\bibfnamefont {S.~A.}\ \bibnamefont
  {Wolf}},\ }\href {\doibase 10.1016/0378-4363(85)90466-8} {\bibfield
  {journal} {\bibinfo  {journal} {Physica B+C}\ }\textbf {\bibinfo {volume}
  {1}},\ \bibinfo {pages} {198} (\bibinfo {year} {1985})}\BibitemShut {NoStop}%
\bibitem [{\citenamefont {Junod}\ \emph {et~al.}(2001)\citenamefont {Junod},
  \citenamefont {Wang}, \citenamefont {Bouquet},\ and\ \citenamefont
  {Toulemonde}}]{junod2001specific}%
  \BibitemOpen
  \bibfield  {author} {\bibinfo {author} {\bibfnamefont {A.}~\bibnamefont
  {Junod}}, \bibinfo {author} {\bibfnamefont {Y.}~\bibnamefont {Wang}},
  \bibinfo {author} {\bibfnamefont {F.}~\bibnamefont {Bouquet}}, \ and\
  \bibinfo {author} {\bibfnamefont {P.}~\bibnamefont {Toulemonde}},\
  }\href@noop {} {\bibfield  {journal} {\bibinfo  {journal} {arXiv preprint
  cond-mat/0106394}\ } (\bibinfo {year} {2001})}\BibitemShut {NoStop}%
\bibitem [{\citenamefont {Margine}\ and\ \citenamefont
  {Giustino}(2013)}]{Margine2013}%
  \BibitemOpen
  \bibfield  {author} {\bibinfo {author} {\bibfnamefont {E.~R.}\ \bibnamefont
  {Margine}}\ and\ \bibinfo {author} {\bibfnamefont {F.}~\bibnamefont
  {Giustino}},\ }\href {\doibase 10.1103/PhysRevB.87.024505} {\bibfield
  {journal} {\bibinfo  {journal} {Phys. Rev. B}\ }\textbf {\bibinfo {volume}
  {87}},\ \bibinfo {pages} {024505} (\bibinfo {year} {2013})}\BibitemShut
  {NoStop}%
\bibitem [{\citenamefont {{Duan}}\ \emph {et~al.}(2014)\citenamefont {{Duan}},
  \citenamefont {{Liu}}, \citenamefont {{Tian}}, \citenamefont {{Li}},
  \citenamefont {{Huang}}, \citenamefont {{Zhao}}, \citenamefont {{Yu}},
  \citenamefont {{Liu}}, \citenamefont {{Tian}},\ and\ \citenamefont
  {{Cui}}}]{Duan14}%
  \BibitemOpen
  \bibfield  {author} {\bibinfo {author} {\bibfnamefont {D.}~\bibnamefont
  {{Duan}}}, \bibinfo {author} {\bibfnamefont {Y.}~\bibnamefont {{Liu}}},
  \bibinfo {author} {\bibfnamefont {F.}~\bibnamefont {{Tian}}}, \bibinfo
  {author} {\bibfnamefont {D.}~\bibnamefont {{Li}}}, \bibinfo {author}
  {\bibfnamefont {X.}~\bibnamefont {{Huang}}}, \bibinfo {author} {\bibfnamefont
  {Z.}~\bibnamefont {{Zhao}}}, \bibinfo {author} {\bibfnamefont
  {H.}~\bibnamefont {{Yu}}}, \bibinfo {author} {\bibfnamefont {B.}~\bibnamefont
  {{Liu}}}, \bibinfo {author} {\bibfnamefont {W.}~\bibnamefont {{Tian}}}, \
  and\ \bibinfo {author} {\bibfnamefont {T.}~\bibnamefont {{Cui}}},\ }\href
  {\doibase 10.1038/srep06968} {\bibfield  {journal} {\bibinfo  {journal} {Sci.
  Rep.}\ }\textbf {\bibinfo {volume} {4}},\ \bibinfo {eid} {6968} (\bibinfo
  {year} {2014})}\BibitemShut {NoStop}%
\bibitem [{\citenamefont {Komelj}\ and\ \citenamefont
  {Krakauer}(2015)}]{Krakauer2015}%
  \BibitemOpen
  \bibfield  {author} {\bibinfo {author} {\bibfnamefont {M.}~\bibnamefont
  {Komelj}}\ and\ \bibinfo {author} {\bibfnamefont {H.}~\bibnamefont
  {Krakauer}},\ }\href {\doibase 10.1103/PhysRevB.92.205125} {\bibfield
  {journal} {\bibinfo  {journal} {Phys. Rev. B}\ }\textbf {\bibinfo {volume}
  {92}},\ \bibinfo {pages} {205125} (\bibinfo {year} {2015})}\BibitemShut
  {NoStop}%
\end{thebibliography}%

\end{document}